\documentclass[conference]{IEEEtran}
\IEEEoverridecommandlockouts
\usepackage[
  backend=biber,
  maxcitenames=2,
  minbibnames=3,
  mincitenames=1,
  maxbibnames=3,
  uniquelist=false,
  style=ieee,
  citestyle=numeric-comp,
  url=false,
  doi=true,
  eprint=true,
  isbn=false,
]{biblatex}
\AtBeginBibliography{\footnotesize}
\usepackage{amsmath,amssymb,amsfonts}
\usepackage{dsfont}
\usepackage{algorithmic}
\usepackage{graphicx}
\usepackage{textcomp}
\usepackage{xcolor}
\usepackage{yquant}
\usepackage{amsthm}
\usepackage{comment}
\usepackage{makecell}
\usepackage{nicefrac}

\usepackage{tabularray}
\UseTblrLibrary{booktabs}
\usepackage{listings} 
\usepackage[inline]{enumitem}
\usepackage[colorlinks=true,urlcolor=teal,
  linkcolor=teal,citecolor=teal]{hyperref} 
\usepackage{glossaries-extra}
\usepackage{tikz}   
\usetikzlibrary{quantikz2,positioning,shadows,shapes.geometric,decorations.pathreplacing}
\usepackage[T1]{fontenc}    
\usepackage{csquotes}
\usepackage{xspace}
\usepackage{censor}
\usepackage{orcidlink}
\usepackage{subcaption}
\usepackage{pgfplots}
\usepackage{numprint}

\newboolean{anonymous}
\setboolean{anonymous}{false}
\setabbreviationstyle[acronym]{long-short}
\glssetcategoryattribute{acronym}{nohyperfirst}{true}

\newacronym{nisq}{NISQ}{noisy intermediate-scale quantum}
\newacronym{qaoa}{QAOA}{quantum approximate optimisation algorithm}
\newacronym{ftqc}{FTQC}{fault tolerant quantum computer}
\newacronym{qml}{QML}{quantum machine learning}
\newacronym{qfm}{QFM}{quantum Fourier model}
\newacronym{vqe}{VQE}{variational quantum eigensolver}
\newacronym{pqc}{PQC}{parameterised quantum circuit}
\newacronym{ml}{ML}{machine learning}
\newacronym{kl}{KL}{Kullback-Leibler}
\newacronym{fft}{FFT}{fast Fourier transform}
\newacronym{rff}{RFF}{random Fourier feature}
\newacronym{mse}{MSE}{mean squared error}
\newacronym{mw}{MW}{Meyer-Wallach}
\newacronym{ef}{EF}{entanglement of formation}
\newacronym{ce}{CE}{concentratable entanglement}
\newacronym{bm}{BM}{Bell measurement}
\newacronym{ree}{REE}{relative entropy of entanglement}
\newacronym{hea}{HEA}{hardware-efficient ansatz}
\newacronym{sea}{SEA}{strongly-entangling ansatz}
\newacronym{bp}{BP}{barren-plateau}
\newacronym{qpu}{QPU}{quantum processing unit}
\newacronym{qoc}{QOC}{quantum optimal control}
\newacronym{fcc}{FCC}{Fourier coefficient correlation}
\newacronym{ode}{ODE}{ordinary differential equation}
\newacronym{drag}{DRAG}{derivative removal by adiabatic gate}
\newacronym{dac}{DAC}{digital analog converter}
\newacronym{rwa}{RWA}{rotating-wave approximation}

\newcommand{\nf}{$^{+}$\xspace}
\newcommand{\uf}{$^{(+)}$\xspace}
\newcommand{\as}{ansatz\xspace}
\newcommand{\ase}{ansätze\xspace}
\newcommand{\As}{Ansatz\xspace}
\newcommand{\Ase}{Ansätze\xspace}

\newcommand{\eg}{\emph{e.g.}\xspace}
\newcommand{\ie}{\emph{i.e.}\xspace}
\newcommand{\cf}{\emph{cf.}\xspace}
\newcommand{\st}{\emph{s.t.}\xspace}

\newcommand{\ketn}[1]{\vert#1\rangle^{\otimes{n}}}
\newcommand{\bran}[1]{\langle#1\vert^{\otimes{n}}}
\newcommand{\Tr}{\operatorname{Tr}}
\newcommand{\btheta}{{\boldsymbol{\theta}}}
\newcommand{\bx}{{\boldsymbol{x}}}
\newcommand{\bp}{{\boldsymbol{p}}}
\newcommand{\bomega}{{\boldsymbol{\omega}}}
\newcommand{\bOmega}{{\boldsymbol{\Omega}}}
\newcommand{\I}{\ensuremath \imath} 
\newcommand{\be}{\ensuremath \boldsymbol{e}}
\newcommand{\fcc}{\ensuremath \mathtt{FCC}_{\Theta}}
\newcommand{\dysontime}{\mathcal{T}}

\DeclareMathOperator*{\argmin}{arg\,min}

\newcommand{\X}{\hat{\mathrm{X}}}
\newcommand{\Y}{\hat{\mathrm{Y}}}
\newcommand{\Z}{\hat{\mathrm{Z}}}
\newcommand{\Rot}{\mathrm{Rot}}
\newcommand{\RX}{\mathrm{\hat{R}_{X}}}
\newcommand{\RY}{\mathrm{\hat{R}_{Y}}}
\newcommand{\RZ}{\mathrm{\hat{R}_{Z}}}
\newcommand{\CX}{\mathrm{C}\X}
\newcommand{\CY}{\mathrm{C}\Y}
\newcommand{\CZ}{\mathrm{C}\Z}
\newcommand{\CRX}{\mathrm{C}\RX}
\newcommand{\CRY}{\mathrm{C}\RY}
\newcommand{\CRZ}{\mathrm{C}\RZ}
\newcommand{\hadamard}{\hat{\mathrm{H}}}
\newcommand{\hamiltonian}{\hat{\mathcal{H}}}
\newcommand{\tone}{$\mathrm{T1}$\xspace}
\newcommand{\ttwo}{$\mathrm{T2}$\xspace}

\makeatletter
\newcommand{\linebreakand}{%
    \end{@IEEEauthorhalign}
    \hfill\mbox{}\par
    \mbox{}\hfill\begin{@IEEEauthorhalign}
}
\makeatother

\newlist{todolist}{itemize}{2}
\setlist[todolist]{label=$\square$}
\usepackage{pifont}

\ifbool{anonymous}{}{}
\ifbool{anonymous}{}{\renewcommand{\blackout}[1]{#1}}
\ifbool{anonymous}{}{}
\ifbool{anonymous}{\newcommand{\genpartemail}[2]{\href{mailto:xxx.xxx@xx.xx}{\blackout{#1}}}}{\newcommand{\genpartemail}[2]{\href{mailto:#1@#2}{#1}}}
\ifbool{anonymous}{\newcommand{\geneurl}[1]{\href{https://thiscatexists.com/}{\blackout{#1}}}}{\newcommand{\geneurl}[1]{\url{#1}}}
\ifbool{false}{\newcommand{\qmlessentials}{\blackout{QML Essentials}\xspace}}{\newcommand{\qmlessentials}{\textsc{QML-Essentials}\xspace}}
\ifbool{anonymous}{\renewcommand{\orcidlink}[1]{}}{}

\definecolor{color1}{HTML}{009682}
\definecolor{color2}{HTML}{DF9B1B}
\definecolor{color3}{HTML}{23A1E0}
\definecolor{color4}{HTML}{002D4C}

\definecolor{lfd1}{HTML}{000000}
\definecolor{lfd2}{HTML}{E69F00}
\definecolor{lfd3}{HTML}{999999}
\definecolor{lfd4}{HTML}{009371}

\definecolor{lfd5}{HTML}{BEAED4}
\definecolor{lfd6}{HTML}{ED665A}
\definecolor{lfd7}{HTML}{1F78B4}

\definecolor{model_col}{HTML}{E69F00}
\definecolor{execution_col}{HTML}{ED665A}
\definecolor{metrics_col}{HTML}{009371}
\definecolor{ansatz_col}{HTML}{E69F00}

\DeclareFixedFont{\ttb}{T1}{txtt}{bx}{n}{6} 
\DeclareFixedFont{\ttm}{T1}{txtt}{m}{n}{6}  
\DeclareFixedFont{\ttmnormal}{T1}{txtt}{m}{n}{9}  
\DeclareFixedFont{\ttmtable}{T1}{txtt}{m}{n}{8}  
\DeclareFixedFont{\ttmfigure}{T1}{txtt}{m}{n}{7}  
\DeclareFixedFont{\ttmsmall}{T1}{txtt}{m}{n}{6}  
\newcommand\pythonstyle{\lstset{
        language=Python,
        basicstyle=\ttm,
        morekeywords=[1]{self, True, False},
        morekeywords=[2]{n_samples,seed,n_qubits,n_layers,circuit_type,initialization,use_multithreading,data_reuploading,n_bins,inputs,execution_type,params,pulse_params},
        keywordstyle=\ttb\color{lfd7},
        keywordstyle=[1]\ttb\color{lfd1},
        keywordstyle=[2]\ttb\color{lfd3},
        emph={Entanglement,Expressibility,Model,Coefficients,FCC},          
        commentstyle=\color{lfd3},
        emphstyle=\ttb\color{lfd2},    
        stringstyle=\color{lfd1},
        showstringspaces=false,
        aboveskip=-8pt,
        belowskip=-5pt,
    }}

\newcommand\pythonstyletable{\lstset{
        language=Python,
        basicstyle=\ttmnormal,
        frame=tb,                         
        showstringspaces=false
    }}
\newcommand\pythoninline[1]{{\pythonstyletable\lstinline!#1!}}
\newcommand\pythoninlinefigure[1]{{\pythonstyletable\lstinline[basicstyle=\ttmfigure]!#1!}}
\newcommand\pythoninlinetable[1]{{\pythonstyletable\lstinline[basicstyle=\ttmtable]!#1!}}

\lstnewenvironment{python}[1][]
{
    \pythonstyle
    \lstset{
        #1
    }
}{}

\pgfdeclarelayer{bg}
\pgfsetlayers{bg,main}

\newlength{\WIDTH}\newlength{\HEIGHT}
\def\cornerrad{3pt}

\tikzset{modulearr/.style={-{Triangle[length=0.5em, scale width=0.7]}, draw=lfd1, line width=1pt}}
\tikzset{doublemodulearr/.style={{Triangle[length=0.5em, scale width=0.7]}-{Triangle[length=0.5em, scale width=0.7]}, draw=lfd1, line width=1pt}}
\tikzset{g/.style={rounded corners=\cornerrad}}

\setlist[enumerate]{label=({\arabic*})}

\begin{document}

\title{Software Between Quantum and Machine Learning -- And Down to Pulses}

\author{
  \IEEEauthorblockN{%
    \blackout{Maja Franz}\textsuperscript{*}\IEEEauthorrefmark{4}\orcidlink{0000-0002-2801-7192},
    \blackout{Melvin Strobl}\textsuperscript{*}\IEEEauthorrefmark{2}\orcidlink{0000-0003-0229-9897},
    \blackout{Jonathan Hunz}\IEEEauthorrefmark{2}\orcidlink{0009-0007-1106-8640},
    \blackout{Lukas Scheller}\IEEEauthorrefmark{2}\orcidlink{0009-0003-9156-7781},\\
    \blackout{Lucas van der Horst}\IEEEauthorrefmark{2}\orcidlink{0009-0003-0609-9582},
    \blackout{Eileen Kuehn}\IEEEauthorrefmark{2}\orcidlink{0000-0002-8034-8837},
    \blackout{Achim Streit}\IEEEauthorrefmark{2}\orcidlink{0000-0002-5065-469X},
    \blackout{Wolfgang Mauerer}\IEEEauthorrefmark{4}\IEEEauthorrefmark{5}\orcidlink{0000-0002-9765-8313}
  }

  \IEEEauthorblockA{
    \IEEEauthorrefmark{4}
    \blackout{Technical University of Applied Sciences Regensburg, Germany},
    \{\genpartemail{maja.franz}{othr.de}, \genpartemail{wolfgang.mauerer}{othr.de}\}@\blackout{othr.de}
  }
  \IEEEauthorblockA{
    \IEEEauthorrefmark{2}
    \blackout{Karlsruhe Institute of Technology, Germany},
    \{\genpartemail{melvin.strobl}{kit.edu}, \genpartemail{lukas.scheller}{kit.edu}, \genpartemail{eileen.kuehn}{kit.edu}, \genpartemail{achim.streit}{kit.edu}\}@\blackout{kit.edu}
  }
  \IEEEauthorblockA{
    \IEEEauthorrefmark{5}
    \blackout{Siemens AG, Technology, Munich, Germany}
  }
}

\maketitle
\begingroup\renewcommand\thefootnote{*}
\footnotetext{Equal contribution}
\endgroup

\begin{abstract}
  Contemporary quantum computing platforms remain, in essence, programmable physical systems whose control is typically mediated through unitary gate abstractions. While such abstractions provide a uniform interface, they obscure important aspects of the underlying hardware and may limit the exploitation of its full capabilities. Direct operation at the level of control pulses offers a more expressive and physically faithful paradigm, enabling, for instance, the implementation of tailored error-mitigation and optimisation strategies. However, this increased expressivity comes at the cost of  greater complexity from the perspective of quantum software development,  necessitating structured and accessible tooling.

  We present a software framework, integrated within the \qmlessentials package, that extends \gls{qml} methodologies to encompass pulse-level modelling. 
  By embedding \gls{qoc} techniques within a \gls{qml} setting, our approach enables the seamless combination of gate-based and pulse-level representations in a unified modelling paradigm.
  The framework provides a comprehensive suite of modelling and analytical capabilities. 
  In particular, we introduce composable ansatz constructions based on interchangeable building blocks, alongside support for end-to-end optimisation of pulse parameters. 
  Motivated by the central role of \glspl{qfm}, we further incorporate a range of Fourier-analytic diagnostics, including the Fourier fingerprint and the \gls{fcc}, complemented by extended measures of entanglement. 
  All performance-critical components are implemented in a high-performance environment using JAX and supported by a dedicated  quantum simulator.
  Taken together, the framework facilitates reproducible and systematic investigations in \gls{qml} research, while bridging the conceptual and practical divide between abstract circuit-based models and hardware-aware pulse-level optimisation. 
  It provides a robust foundation for future developments at the intersection of quantum machine learning and quantum control.
\end{abstract}


\glsresetall

\section{Introduction}

\Glspl{pqc} constitute the core of contemporary \gls{qml}, serving as function approximators across a variety of supervised and unsupervised \gls{ml} tasks~\cite{schuld_supervised_2018,franz_uncovering_2022,yue_challenges_2023,heimann_learning_2024,carbonelli_challenges_2024,eisert_mind_2025,huang_generative_2025}.
A Fourier-analytic perspective~\cite{prezsalinas_data_2020,schuld_effect_2021} allows studying their expressive power (\ie, the classes of functions that can be learned). 

By characterising \glspl{pqc} and their data encodings through their induced frequency spectra, \emph{\glspl{qfm}} provide a structured approach to understanding how quantum circuits realise function classes in terms of Fourier coefficients and frequencies.
At the same time, a growing body of literature highlights the fundamental limitations in \glspl{pqc}.
For instance, \emph{barren plateaus}~\cite{wang_noise_2021,larocca_barren_2025}, or trainability degradation~\cite{depalma_limitations_2023,ahmed_comparative_2025} and restricted effective expressivity~\cite{fontana_spectral_2022,franz_out_2025} under realistic hardware constraints restrict the capabilities of quantum computing approaches in \gls{ml}.
Despite this increasing number of hardware-aware studies, we see a need for systematic tools that enable investigations beyond idealised quantum gates.
In particular, while most \gls{qml} studies assume unitary gate decompositions~\cite{cerezo_challenges_2022,bharti_noisy_2022}, \glspl{qpu}, such as superconducting systems, operate at the level of control pulses.
However, based on Refs.~\cite{krantz_quantum_2019,meitei_gate_2021,liang_variational_2022,melo_pulse_2023,tao_role_2024,sarma_designing_2025,tao_design_2025}, we see a large potential in advancing \gls{qml} research by directly learning pulse-level control sequences instead of gate angles.

We therefore aim to bridge the gap between the \emph{high-level} circuit analysis metrics and the \emph{low-level} pulse-level control with a software framework that
\begin{enumerate*}[label=(\arabic*)]
  \item provides a modular and systematically composable \gls{qml} model construction,
  \item integrates rigorous diagnostic tools tailored to the Fourier-analytic perspective of \gls{qml}, and
  \item supports pulse-level modelling grounded in \gls{qoc}.
\end{enumerate*}
Rather than treating control pulses as an implementation detail, we expose them as differentiable objects, enabling end-to-end optimisation of pulse parameters.
This enables the study of expressivity, trainability and other circuit properties, such as the entangling capability, directly at the level of physically realisable controls.
Additionally, pulse-level learning paves the way for certain error mitigation techniques~\cite{henao_adaptive_2023}, where control signals are optimised to suppress coherent errors or reduce sensitivity to noise~\cite{greiwe_effects_2023,thelen_approximating_2024,maschek_make_2025}.

In this work, we introduce software tools in Python that provide a modular approach to variational \as design that can be instantiated by conventional unitary gates or as implicit pulse-level control sequences.
Using this dual representation allows controlled comparisons between circuit-level models and hardware-aware implementations.
Specifically, we extend the \qmlessentials package, introduced in Ref.~\cite{strobl_qml_2025}, which already delivers some tools and metrics for circuit analysis from a Fourier-analytic perspective. 

Building on \glspl{qfm} as a central theme, we incorporate additional circuit metrics such as the Fourier fingerprint or the \gls{fcc}~\cite{strobl_fourier_2025}.
Furthermore, we expand the set of entangling capability measures by incorporating \gls{ef} and \gls{ce}, which are also applicable to mixed (noisy) quantum states.

To summarise, our software tools in \qmlessentials provide a comprehensive, reproducible~\cite{mauerer_1_2022} analysis and benchmarking suite for studying the relationship between entanglement, model expressivity, and efficient, hardware-agnostic circuit representations.
Additionally, all performance-critical components are implemented with JAX~\cite{bradbury_jax_2024} and supported by a dedicated and custom differentiable simulator for fast array computations.

The remainder of this article is structured as follows:
\autoref{sec:relwork} reviews the state of the art in \gls{qml}-software with a focus on \glspl{qfm} and pulse-level learning, followed by a brief introduction to \glspl{qfm} in \autoref{sec:preliminaries}.
In \autoref{sec:package-overview}, we then present an overview of our software and its integration in the \qmlessentials package from Ref.~\cite{strobl_qml_2025}.
Throughout this article, novel features compared to \qmlessentials are marked with \nf, and major changes with \uf.
A major part of our software is the tooling for pulse-level learning, which we present in \autoref{sec:pulses}.
Subsequently, in \autoref{sec:metrics} we describe our implemented \as diagnostics metrics for the spectral properties, expressibility and entangling capability with some proof-of-concept results.
Finally, we give an outlook on the possibilities of future work and experiments that utilise our toolkit and conclude in \autoref{sec:concl}.

\section{Related Work}
\label{sec:relwork}

The current state of the art in \gls{qml} software can be roughly separated into three main categories, namely
\begin{enumerate*}[label=(\arabic*)]
  \item general libraries for quantum circuit simulation or real hardware execution, at the core of established classical machine learning frameworks,
  \item \gls{qml}-tailored frameworks, which, in addition to quantum circuit execution, offer useful tools for \gls{ml}-research, such as predefined datasets and differentiable programming, or
  \item individual code setups accompanying research articles for specific tasks or reproduction, often leveraging tools from the first two categories.
\end{enumerate*}

General quantum frameworks, such as Qiskit~\cite{javadiabhari_quantum_2024}, t$\ket{\text{ket}}$~\cite{sivarajah_t_2020}, or Cirq~\cite{developers_cirq_2025}, which offer comprehensive support for transpilation, real hardware execution, and stable circuit simulations, fall into the first category.
These frameworks provide a variety of simulation, execution and analysis capabilities, and even some quantum algorithmic primitives, like Trotterisation and parameter optimisation functionality for the \gls{qaoa}, or \gls{vqe}.
Furthermore, SimuQ~\cite{peng_simuq_2024}, and QuTiP~\cite{li_pulse_2022} support extended functionality for analysing quantum computers at pulse-level.
Here, SimuQ focuses on Hamiltonian simulation, and QuTiP provides several algorithms for \gls{qoc}.

However, customisation and effort are required to utilise quantum circuits defined in one of those frameworks in a \gls{qml} setting.

The second category contains more \gls{qml}-tailored approaches, including frameworks like PennyLane~\cite{bergholm_pennylane_2022}, TensorFlow Quantum~\cite{broughton_tensorflow_2021}, or Qibo~\cite{efthymiou_qibo_2022} with its recent \gls{qml}-library~\cite{robbiati_qiboml_2025}.
They provide a comprehensive toolset that extends the capabilities of the quantum circuit simulation or hardware interface with, for instance, quantum and classical \gls{ml} datasets, different gradient computation methods, or a native integration into classical \gls{ml} frameworks, like PyTorch~\cite{paszke_pytorch_2019}, or TensorFlow~\cite{abadi_tensorflow_2015}.
The recent approach AutoQML~\cite{roth_autoqml_2025} bundles the functionality of these \gls{qml} frameworks to generate automated pipelines for non-domain experts and industrial use cases, as a more specialised method compared to, for instance, Refs.~\cite{fingar_quark_2022,gierisch_qef_2025}, which orchestrate general quantum experiments.
The two frameworks, PennyLane and Qibo, which are most similar to our work, additionally integrate differentiable pulse-level control within a \gls{qml} workflow, enabling a full-stack approach from application to hardware or simulator execution.
Our work differs from these two in that we particularly focus on the analysis of the interplay between \glspl{qfm} and their properties.

Since it has been established that \glspl{pqc} can be represented as truncated Fourier series~\cite{prezsalinas_data_2020,schuld_circuit_2020} (\cf \autoref{sec:preliminaries}), the Fourier spectrum poses an important tool to determine the expressivity of a \gls{qml} model.
From a fundamental research perspective, some relations between other properties of quantum circuits and their Fourier spectrum are known, such as the expressibility~\cite{sim_expressibility_2019,mhiri_constrained_2024} (\ie, the ability of a model to explore the Hilbert space; \cf \autoref{sec:expressibility}), or dequantisability~\cite{schreiber_classical_2023,sweke_potential_2025} (\ie, the ability to find classical alternatives to represent a quantum model).
Furthermore, a variety of publications~\cite{fontana_spectral_2022,casas_multidimensional_2023,nemkov_fourier_2023,jaderberg_let_2024,wiedmann_fourier_2024,strobl_fourier_2025,franz_out_2025,duffy_spectral_2026} investigate the Fourier spectrum to improve \gls{qml} algorithms, for instance, to find suitable circuit structures or data encodings~\cite{gogeissl_quantum_2024}.
However, if code accompanies these kinds of research articles, it mainly consists of individual code setups, which may allow reproducing individual numerical experiments, yet are difficult to build upon.
Therefore, these software packages fall into our third category of \gls{qml} software.

The \qmlessentials package, introduced in Ref.~\cite{strobl_qml_2025}, aims to close the gap between the categories (2) and (3):
It provides a set of standard tools, less general than category (2) frameworks, but tailored to \glspl{qfm} investigations and suited as a backend for category (3) numerical experiments.
However, we find that the unitary-centred approach in Ref.~\cite{strobl_qml_2025} masks the possibilities of lower-level quantum computations in \gls{qml} to directly access and learn the pulses of an operation.
With our \gls{qoc} implementation, we aim to create a differentiable interface for gradient-based pulse optimisation, such as algorithms suggested in Refs.~\cite{khaneja_optimal_2005,machnes_tunable_2018,morzhin_krotov_2019}, or more recent approaches that, for instance, use reinforcement learning~\cite{sarma_designing_2025}.

Additionally, we propose further improvements to \qmlessentials~\cite{strobl_qml_2025} to mature them into a software framework that comprehensively tackles \glspl{qfm}. 

All performance-critical computations (including all quantum computational simulations) are implemented using the JAX~\cite{bradbury_jax_2024} library, and a variety of other features (\cf \autoref{sec:package-overview}).

\begin{figure*}[t]
  \input{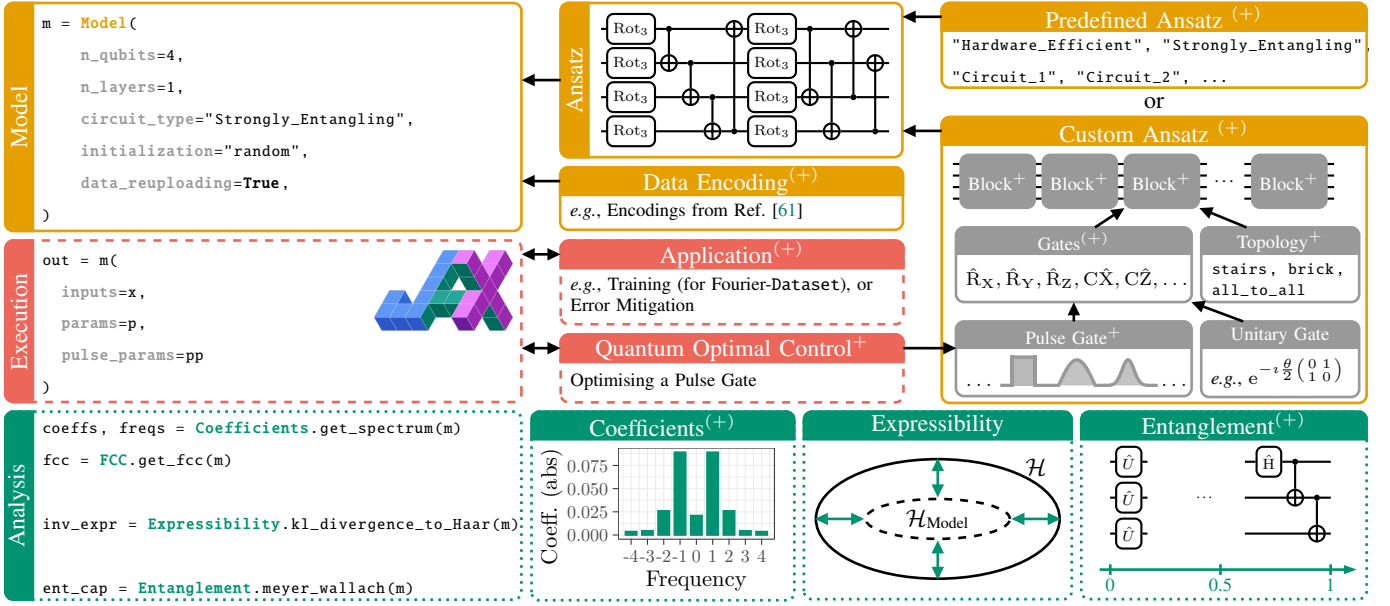}

  \caption{%
    Overview of our contributions to \qmlessentials with \nf indicating novel additions, and \uf major changes.
    The code example in Python on the left-hand side describes the main functionality of the framework.
    The \textcolor{model_col}{solid yellow} blocks on the right hand show the possibilities for defining a quantum Fourier \textcolor{model_col}{model}, based on different \textcolor{ansatz_col}{\ase} and \textcolor{ansatz_col}{encodings}.
    \textcolor{execution_col}{Dashed red} blocks show the possibilities of utilising the \textcolor{execution_col}{execution} of a model, which encompasses the JAX-based~\cite{bradbury_jax_2024} classical simulation of a quantum circuit and its measurement.
    Different input, parameter and pulse parameter values open up possibilities for a variety of \textcolor{execution_col}{applications} and \textcolor{execution_col}{QOC}.
    \textcolor{metrics_col}{Dotted green} blocks correspond to built-in metrics for analysing the model, including computing the \textcolor{metrics_col}{frequencies and coefficients} and \textcolor{metrics_col}{FCC} of the corresponding Fourier spectrum, as well as different metrics for \textcolor{metrics_col}{expressibility} and the \textcolor{metrics_col}{entangling capability}.
  }
  \label{fig:software_overview}
\end{figure*}

\section{Preliminaries on Quantum Fourier Models}
\label{sec:preliminaries}

The goal of \gls{qml} aligns with classical machine learning: \emph{learning} a function $f$ using a function approximator $\tilde{f}$ parametrised by $\btheta$.
The parameters $\btheta$ are \emph{trained} to minimise the difference between $f(\bx)$ and $\tilde{f}(\bx, \btheta)$ according to some metric, for a $D$-dimensional input $\bx = (x_1, \dots, x_D) \in \mathbb{R}^D$, using classical optimisation routines, such as gradient descent~\cite{schuld_supervised_2018}.

In \gls{qml}, the function approximator is an $n$-qubit quantum circuit whose expectation value, using the observable $\mathcal{M}$, is given by
\begin{equation}
  \tilde{f}(\bx, \btheta)=\bran{0} \hat{U}^{\dagger}(\bx, \btheta) \mathcal{M} \hat{U}(\bx, \btheta)\ketn{0},
  \label{eq:circuitExpr}
\end{equation}
parametrised by the trainable parameter vector $\btheta = (\btheta_1, \dots, \btheta_{L+1}) \in [\mathbb{R}^K]^{L+1}$ with $L$ layers of $K$ parameters per layer\footnote{Without loss of generality, we assume identical numbers of parametrised gates per layer.}.

We build the eventual unitary $\hat{U}(\bx, \btheta)$ as products of multiple layers ($\ell \in [1,L]$) of consecutive encoding unitaries $\hat{S}^{(\ell)}(\bx)$ and trainable unitaries $\hat{W}^{(\ell)} \coloneqq \hat{W}^{(\ell)}(\btheta_\ell)$ as follows:
\begin{equation}
  \hat{U}(\bx)=\hat{W}^{(L+1)} \hat{S}^{(L)}(\bx) \hat{W}^{(L)} \cdots \hat{W}^{(2)} \hat{S}^{(1)}(\bx) \hat{W}^{(1)}
\end{equation}
where an additional last trainable layer $\hat{W}^{(L+1)}$ is added.
To allow for an arbitrary circuit structure and placement of unitary and encoding gates, we construct $\hat{S}^{(\ell)}(\bx)$ to comprise combinations of $D$ unitaries $\hat{S}^{(\ell)}_i(\bx)$, each of which encodes one single input feature $x_i$:
\begin{equation}
  \hat{S}_i^{(\ell)}(\bx) = \exp{(-\I \be_i^T \bx \hat{G}_i^{(\ell)})} = \exp{(-\I x_i \hat{G}_i^{(\ell)})}
  \label{eq:datareup}
\end{equation}
with Hermitian generator $\hat{G}_i^{(\ell)}$ and standard basis vector $\be_i$.

As established in Refs.~\cite{prezsalinas_data_2020,schuld_effect_2021}, this architecture allows rewriting~\autoref{eq:circuitExpr} as a partial Fourier series
\begin{equation}
  \tilde{f}(\bx, \btheta) = \sum_{\bomega \in \bOmega} c_{\bomega}(\btheta) e^{\I \bomega^T \bx},
  \label{eq:qfm}
\end{equation}
where $\bOmega$ contains the unique frequencies resulting from the eigenvalues of each generating encoding Hamiltonian $\hat{G}_i^{(\ell)}$.
The corresponding complex Fourier coefficients are given by $\{c_{\bomega}(\btheta)\}_{\bomega \in \bOmega}$.

From \autoref{eq:qfm} it can be inferred that the Fourier frequencies derive from the encoding gates (\ie, the feature map), while the Fourier coefficients, parametrised by the parameters $\btheta$ of the circuit, solely depend on the circuit structure, which we refer to as \emph{\as} in the further course of this work.
In \autoref{sec:fourier-metrics}, we describe our implemented metrics for circuit analysis that derive from the Fourier spectrum and how we compute coefficients in the framework.

\section{Package Overview}
\label{sec:package-overview}

An overview of the different components included in \qmlessentials and a code example are shown in \autoref{fig:software_overview}.
At a more technical level, the Python modules of the package are summarised in \autoref{tab:modules}.
In the following, we give a more detailed description of the modules and components utilised in this work, while a full API documentation is delivered with the software package.
The source code is hosted on GitHub\footnote{\geneurl{https://github.com/cirKITers/qml-essentials}}, linked to Zenodo~\cite{franz_qml_0000}, and distributed via PyPI\footnote{\geneurl{https://pypi.org/project/qml-essentials/}}.

\begin{table}[]
  \centering
  \caption{Python module overview.}
  \label{tab:modules}
  \begin{tblr}{width=\linewidth,
    colspec={lX[l]},
    row{2-Z}={belowsep=0.15em}}
    \toprule
    \textbf{Module}                                                        & \textbf{Description}                                                                                                                                           \\
    \midrule
    \pythoninlinetable{model.py}\uf                                        & Data-reuploading model class with various options to parameter initialisation, encoding schemes, measurement type and noise.                                   \\

    \pythoninlinetable{ansaetze.py}\uf                                     & Predefined \ase and encodings to be used within a model.            \\
    \begin{tabular}[t]{@{}l@{}}
      \pythoninlinetable{gates.py}\uf,
      \pythoninlinetable{unitary.py}, \\
      \pythoninlinetable{pulses.py}\nf
    \end{tabular} & Collection of (noisy) gates implemented as both unitary and pulse variant.                                                                                                                           \\
    \pythoninlinetable{topologies.py}\nf                                   & Collection of circuit topologies.                                                                                                                              \\

    \pythoninlinetable{qoc.py}\nf                                          & Tools for \gls{qoc} and pulse-level based optimisation (\cf \autoref{sec:pulses}).                                                                             \\
    \pythoninlinetable{coefficients.py}\uf                                 & Calculation of Fourier coefficients and frequencies (\gls{fft} or analytical), as well as the \gls{fcc}. Also includes a Fourier series dataset generator. \\
    \pythoninlinetable{expressibility.py}                                  & Calculation of expressibility (\cf \autoref{sec:expressibility} for implemented metrics).                                                                  \\
    \pythoninlinetable{entanglement.py}\uf                                 & Calculation of entangling capability (\cf \autoref{sec:entanglement} for implemented metrics).                                                             \\
    \begin{tabular}[t]{@{}l@{}}
      \pythoninlinetable{tape.py}\nf,
      \pythoninlinetable{script.py}\nf, \\
      \pythoninlinetable{jaqsi.py}\nf
    \end{tabular} & JAX-based quantum circuit simulator.                                                                                                             \\
    \begin{tabular}[t]{@{}l@{}}
      \pythoninlinetable{operations.py}\nf, \\
      \pythoninlinetable{math.py}\uf,
      \pythoninlinetable{utils.py}\uf \\
      \pythoninlinetable{drawing.py}\nf
    \end{tabular} & Collection of utilities in the framework for \eg quantum operations, mathematical tools, or drawing. \\

    \bottomrule
  \end{tblr}
\end{table}

\subsection{Model and \As Overview}

The \pythoninline{Model} forms the core of the package, built upon a given \as.
An \as can be imported from the \as-library, which contains, among others, a \emph{\gls{sea}}~\cite{schuld_circuit_2020} (shown as an example at the top of \autoref{fig:software_overview}), a \emph{\gls{hea}}~\cite{kandala_hardware_2017} and almost all \ase introduced in Ref.~\cite{sim_expressibility_2019}.
Alternatively, we refactored custom \as definitions to allow composition via multiple \pythoninline{Block} instances, as shown on the right-hand side of \autoref{fig:software_overview}.
A \pythoninline{Block} takes a set of quantum gates and a topology, which describes the order and layout of two-qubit operations in a circuit.

The resulting ansatz represents one variational layer of the model of size \pythoninline{n_qubits}, which is repeated multiple times, depending on the \pythoninline{n_layer} argument of the \pythoninline{Model} constructor, to build the full circuit.
For \gls{qml} use cases, the unitary \as comprises multiple rotational gates (\eg, $\RX$, $\RY$, $\RZ$), parametrised by at least one angle. 

We refer to the set of angles corresponding to all rotational gates in the circuit as the \emph{parameters} of the model, which may be subject to optimisation in a training process.
For the initialisation of the parameters, we provide multiple approaches, such as setting the parameters to specific values (zero or $\pi$), or uniformly at random in a specific domain.

Following the data re-uploading mechanism from \autoref{eq:datareup}, we support input encodings interleaved with the variational layers, where encoding positions can be controlled with the \pythoninline{data_reuploading} argument of the \pythoninline{Model} constructor.
While the default input encoding scheme is an angle encoding (using a $\RX$ gate), we also implemented the encoding schemes of Ref.~\cite{peters_generalization_2023}:
As introduced in \autoref{sec:preliminaries}, the number of frequencies (and therefore the number of coefficients) depends on the chosen feature map.
Hence, we provide an interface to choose between different encoding methods in addition to providing a custom feature map:
\begin{enumerate*}[label=(\arabic*)]
  \item Hamming encoding (default, linear spectrum),
  \item binary encoding (exponential spectrum), and
  \item ternary encoding (exponential spectrum).
\end{enumerate*}
In addition, trainable parameters can be applied in the feature map as introduced in Ref.~\cite{jaderberg_let_2024}.

Additionally, we allow to view \ase from a lower abstraction level than unitary gates, by providing the feature to parametrise an \as directly with pulses, using \emph{pulse parameters} (\cf \autoref{sec:pulses}), instead of rotational angles.

\subsection{Execution with Input- and Parameter-Batching}
Depending on the application or metric that is to be evaluated for a model, different modes of execution may be required.
In \qmlessentials, this mode can be selected by setting the \pythoninline{execution_type} argument when calling the model.
In particular, this value can be set to
\begin{enumerate*}[label=(\arabic*)]
  \item \pythoninline{"expval"}, which gives the expectation value of an observable (\eg, the default is $\hat{Z}$, acting on a specified set of qubits),
  \item \pythoninline{"density"}, which gives the density matrix,
  \item \pythoninline{"state"}, which gives the state-vector representation, and
  \item \pythoninline{"probs"}, which gives the output probability distribution over a specified set of qubits, using a finite number of shots.
\end{enumerate*}

To evaluate models with concurrent sets of inputs and (pulse) parameters, we handle batch dimensions internally to cover all possible combinations.

We utilise our own simulator to carry out the computation of the circuit with the specified observables.
This simulator and all implementations of the metrics presented in this work are entirely built upon the JAX~\cite{bradbury_jax_2024} library for high-performance array computations and utilisation of parallelisation where possible.
To solve the differential equation required for pulse-level simulation, our framework supports adaptive Runge-Kutta methods (e.g., Dormand-Prince of order $8$~\cite{dormand_family_1980} via Diffrax~\cite{kidger_neural_2022}) as well as commutator-free Magnus integrators tailored for highly oscillatory Hamiltonian dynamics. 

Our simulator is continuously tested against the PennyLane~\cite{bergholm_pennylane_2022} simulator to ensure correctness of the computations.

\subsection{Other Features}
\label{sec:features}
The support for pulse-parameters and \gls{qoc} is detailed in \autoref{sec:pulses}.
Furthermore, the \qmlessentials package offers circuit analysis metrics for computing the Fourier spectrum, expressibility, or entangling capability.
In \autoref{sec:metrics} we describe further which metrics are part of the original framework and how we contribute towards improving the existing state.
In addition, we added other notable features, which are summarised in this subsection.

\subsubsection{Dataset Generator}
To reduce the overhead required to run simple trainings of \glspl{qfm}, we provide a Fourier series dataset generator.
Given a model, its intrinsic frequencies are inferred based on the encoding strategy chosen upon initialisation.
Afterwards, corresponding coefficients are uniformly sampled within a unit circle.
The resulting Fourier series is then used to obtain pairs of data points ($x$ and $\mathcal{F}(x)$), which can then be used in a training environment.

\subsubsection{Visualisation}
To obtain a visualisation for a model's circuit, our framework offers an export of circuit diagrams as text-based or Matplotlib figures.
Additionally, we support exporting to the Quantikz \LaTeX{} package~\cite{kay_tutorial_2023}.
The resulting \LaTeX{} code can be easily copied into a paper or thesis.
Pulse-sequence schedules can also be visualised via the same API.

\section{Pulse-Level Simulation}
\label{sec:pulses}

Generally, unitary gates are used to train \gls{qfm} models.
However, there exist lower-level parameters that allow controlling the exact pulse of each gate applied to a quantum system, to which we refer as \emph{pulse parameters}.
These parameters are usually subject to a fine-tuning process carried out by the hardware provider in a process commonly termed \gls{qoc}.
In our framework, we provide an interface to conduct experiments that require access to the pulse parameters, therefore enabling research that is more tailored to the actual quantum hardware.

\subsection{Pulses of Basis Gates}
\label{sec:pulse-shapes}

Generally, the dynamics of a closed quantum system are described by the unitary time evolution of a state $\psi$.
For many platforms (\eg, superconducting transmons), dynamics from time $t_0$ to $t_0 + \delta t$ are governed by the time-dependent Schrödinger equation~\cite{efthymiou_qibo_2022,rothesantos_evaluation_2025} (assuming the convention $\hbar=1$):
\begin{equation}
  \ket{\psi(t_0 + \delta t)} = \dysontime \exp\left(-\I \int_{t_0}^{t_0 + \delta t} \hamiltonian(t, \bp)\mathrm{d}t\right) \ket{\psi(t_0)},
  \label{eq:evolution}
\end{equation}
where $\dysontime$ denotes the time-ordering operator, and $\hamiltonian(t, \bp)$ is the time-dependent Hamiltonian parametrised by pulse parameters $\bp$ (distinct from the Hadamard gate $\hadamard$).

Typically, the Hamiltonian decomposes into a static operator $\hamiltonian_\text{static}$ (unperturbed energy levels) and $m$ control operators $\{\hamiltonian_j\}_{j=1}^m$ driven by time-dependent signals $S_j(t; \bp)$, such that
\begin{equation}
  \hamiltonian(t, \bp) = \hamiltonian_\text{static} + \sum_{j=1}^m S_j(t; \bp) \hamiltonian_j.
\end{equation}
Throughout our work, we choose $\hamiltonian_\text{static} = \frac{\omega_{q}}{2}\Z$ for general periodically driven two-level systems~\cite{wu_strong_2007}, with the qubit frequency $\omega_{q}=10\pi$ adapted as needed for a specific hardware.
Notably, both the exact static Hamiltonian and control Hamiltonians depend on the specific hardware that is used~\cite{alexander_qiskit_2020,votto_universal_2024}.
The control signals are typically composed of an envelope function $E_j(t; \bp)$ and a carrier wave: $S_j(t; \bp) = E_j(t; \bp) \cos(\omega_c t + \phi_c)$, where $\omega_c$ is the carrier frequency and $\phi_c$ is the carrier phase.

To simplify the evolution in \autoref{eq:evolution}, it is common to move to the interaction picture with respect to $\hamiltonian_\text{static}$, factoring out the free evolution. 
Our software framework supports the exact numerical integration of these full interaction-picture dynamics, either directly in the laboratory frame or via a product-to-sum drive frame decomposition that explicitly resolves slow and fast oscillatory modes. 
However, for the numerical experiments in this work, we primarily utilise the \gls{rwa}, which drops the fast-oscillating counter-rotating terms. 
The \gls{rwa} significantly reduces the integration stiffness and accelerates simulation while maintaining high accuracy for near-resonant drives.

Selecting the control envelopes $E_j(t; \bp)$ such that the resulting time evolution implements a target unitary operation is then the goal of \gls{qoc}.
To this end, we provide a set of commonly known pulse envelope shapes, listed in \autoref{tab:pulse_envelopes}.

\begin{table}[htb]
  \centering
  \caption{Pulse envelope functions, parametrised by the time $t$ and pulse parameters $\bp$, consisting of combinations of Amplitude $A$, pulse width $\sigma$, the central pulse time $t_c$, and, in case of the \gls{drag} pulse, a scaling factor $\beta$.}
  \label{tab:pulse_envelopes}
  \footnotesize
  \renewcommand{\arraystretch}{0.95}

  \begin{tblr}{
    width=\linewidth,
    colspec={Q[l,wd=32mm] X[l]},
    column{2}={mode=math},
    colsep=2pt,
    leftsep=0pt,
    rightsep=0pt,
  }
    \toprule
    Envelope                                            & E(t; \bp)       \\
    \midrule
    Gaussian                                            &
    A\,\exp\!\left(-\nicefrac12\left(\frac{t-t_c}{\sigma}\right)^2\right) \\
    Rectangle                                           &
    A\,\mathbf{1}\!\left(|t-t_c|\le \nicefrac{\sigma}{2}\right)           \\
    Raised Cosine                                       &
    A\cos\!\left(\pi\frac{t-t_c}{\sigma}\right)\quad
    \forall\,|t-t_c|\le \nicefrac{\sigma}{2}                              \\
    \gls{drag}~\cite{motzoi_simple_2009} &
    A\,\exp\!\left(-\nicefrac12\left(\frac{t-t_c}{\sigma}\right)^2\right)
    \left[1-\I \beta\,\frac{t-t_c}{\sigma^2}\right]                         \\
    Hyperbolic Secant                                   &
    A\left(\cosh\!\left(\frac{t-t_c}{\sigma}\right)\right)^{-1}                    \\
    \bottomrule
  \end{tblr}
\end{table}

\begin{figure}[t]
  \includegraphics[width=\columnwidth]{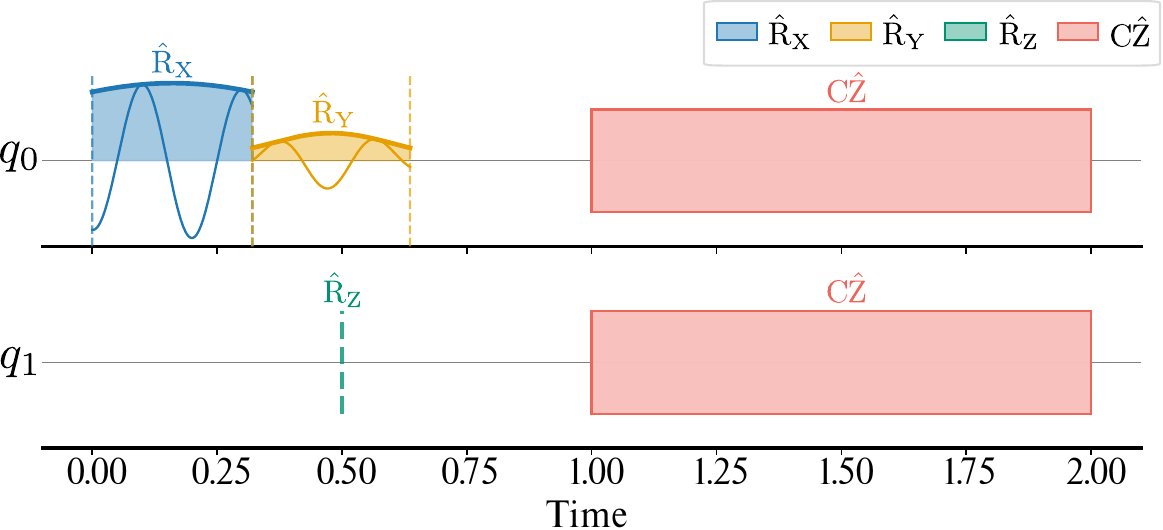}

  \caption{%
    Pulse sequence diagram showing an $\RX$ and $\RY$ gate on the first-, a $\RZ$ on the second, and a $\CZ$ gate on both qubits.
    Shaded areas show envelopes with carrier frequencies marked by thin lines.
    Vertical lines delimit the gate duration.
  }
  \label{fig:pulse_sequence}
\end{figure}

Given the envelope and parameters subject to \gls{qoc} (\cf \autoref{sec:qoc}), a pulse sequence diagram can be used to visualise the envelope, carrier frequency and duration of the pulses corresponding to the particular gates of a given circuit.
An exemplary diagram is shown in \autoref{fig:pulse_sequence} where shaded areas represent the envelope and dashed vertical lines fence the duration of the pulse, in between which the carrier frequency is rendered.
Note that in case of the $\RZ$ and $\CZ$ gates, no actual pulse is applied; instead, the system evolves freely to rotate the phase accordingly.

\subsection{Pulse-Gate Interface}
\label{sec:pulse_gates}

Carefully optimising the fidelities of a set of basis gates allows building composed gates upon this foundation. 

\begin{figure}
  \centering
  \setlength{\WIDTH}{\columnwidth}
\setlength{\HEIGHT}{0.2\textheight}

\def\gatewidth{2.5em}
\def\gateheight{2.25em}
\def\subgateheight{1.5em}
\def\leveldist{2.3em}
\def\gatemargin{4pt}

\tikzset{gate/.style={text=white, font=\small, line width=1pt, minimum width=\gatewidth, minimum height=\subgateheight}}
\tikzset{gateframe/.style={rounded corners=\cornerrad, rectangle, minimum width=\gatewidth, minimum height=\gateheight, line width=1pt}}
\tikzset{emptygate/.style={minimum width=\gatewidth, minimum height=\gateheight}}
\tikzset{param/.style={font=\scriptsize, inner sep=2pt}}
\tikzset{filler/.style={fill=white, minimum width=\gatewidth+\gatemargin, minimum height=\gatewidth+\gatemargin}}
\tikzset{whiteborder/.style={-{Triangle[length=0.3em, scale width=0.2]}, draw=white, line width=3pt}}

\newcommand{\gatewithparams}[6]{%
  \node[#3, emptygate] (#1) #5 {};
  \node[filler] at (#1.center) {};
  \node[#3, draw=#4, gateframe] (#1) #5 {};
  \fill[#4] ($(#1.north east) + (0,-\subgateheight)$) -- ($(#1.north west) + (0,-\subgateheight)$) {[rounded corners=\cornerrad]-- (#1.north west) -- (#1.north east)} -- cycle;
  \node[gate, anchor=north] at (#1.north) {#2};
  \node[param, anchor=south] at (#1.south) {#6};
}

\begin{tikzpicture}
  \coordinate (tl) at (0, \HEIGHT);
  \coordinate (tr) at (\WIDTH, \HEIGHT);
  \coordinate (bl) at (0, 0);
  \coordinate (br) at (\WIDTH, 0);


  \gatewithparams{cry}{$\CRY$}{anchor=north west}{lfd1}{at (tl)}{24}
  \gatewithparams{cy}{$\CY$}{anchor=north east}{lfd1}{at (tr)}{11}
  \gatewithparams{crx}{$\CRX$}{anchor=north, xshift=0.25*\gatewidth}{lfd1}{at ($(tl)!0.33!(tr)$)}{26}
  \gatewithparams{crz}{$\CRZ$}{anchor=north, xshift=-0.25*\gatewidth}{lfd1}{at ($(tl)!0.66!(tr)$)}{20}

  \gatewithparams{cx}{$\CX$}{anchor=north west}{lfd3}{at ($(cry.south west) + (0, -\leveldist)$)}{9}
  \gatewithparams{h}{$\hadamard$}{anchor=north east}{lfd3}{at ($(cy.south east) + (0, -\leveldist)$)}{4}

  \draw[modulearr] (cx) -- (h);
  \gatewithparams{rz}{$\RZ$}{anchor=north west}{lfd4}{at ($(crz.west|-cx.south) + (0, -\leveldist)$)}{1}

  \gatewithparams{cz}{$\CZ$}{anchor=north west}{lfd4}{at ($(cry.west|-cx.south) + (0, -\leveldist)$)}{1}
  \gatewithparams{ry}{$\RY$}{anchor=north west}{lfd4}{at ($(crx.west|-cx.south) + (0, -\leveldist)$)}{3}
  \gatewithparams{rx}{$\RX$}{anchor=north east}{lfd4}{at ($(cy.east|-h.south) + (0, -\leveldist)$)}{3}

  \draw[whiteborder] (h) -- (ry);
  \draw[modulearr, dotted] (h) -- (ry);

  \draw[whiteborder] (cry) -- (cx);
  \draw[modulearr] (cry) -- (cx);

  \draw[whiteborder] (cry) -- (ry);
  \draw[modulearr] (cry) -- (ry);

  \draw[whiteborder] (crx) -- (cx);
  \draw[modulearr] (crx) -- (cx);

  \draw[whiteborder] (crx) -- (rz);
  \draw[modulearr] (crx) -- (rz);

  \draw[whiteborder] (crx) -- (ry);
  \draw[modulearr] (crx) -- (ry);

  \draw[whiteborder] (crz) -- (cx);
  \draw[modulearr] (crz) -- (cx);

  \draw[whiteborder] (cx) -- (cz);
  \draw[modulearr, dotted] (cx) -- (cz);

  \draw[whiteborder] (cx) -- (ry);

  \draw[whiteborder] (cy) -- (rz);
  \draw[modulearr] (cy) -- (rz);

  \draw[whiteborder] (crz) -- (rz);
  \draw[modulearr] (crz) -- (rz);

  \draw[whiteborder] (cy) -- (cx);
  \draw[modulearr, dotted] (cy) -- (cx);

  \draw[whiteborder] (h) -- (rz);
  \draw[modulearr, dotted] (h) -- (rz);

  \coordinate (mid) at ($(crx.south)!0.5!(crz.south)$);
  \gatewithparams{rot3}{$\Rot_3$}{anchor=north}{lfd3}{at ($(mid) + (0, -1.68*\leveldist)$)}{5}
  \draw[whiteborder] (rot3) -- (ry);
  \draw[modulearr, dotted] (rot3) -- (ry);

  \draw[whiteborder] (rot3) -- (rz);
  \draw[modulearr] (rot3) -- (rz);

\end{tikzpicture}
  \caption{Overview of the connectivity between \textcolor{lfd4}{basis gates} and \textcolor{lfd1}{composed gates}. The line type (dotted/solid) represents the number of child gates (1/2) the parent gate depends on. Labels at the bottom of each node represent the total number of pulse parameters including redundancies.}
  \label{fig:gates_graph}
\end{figure}
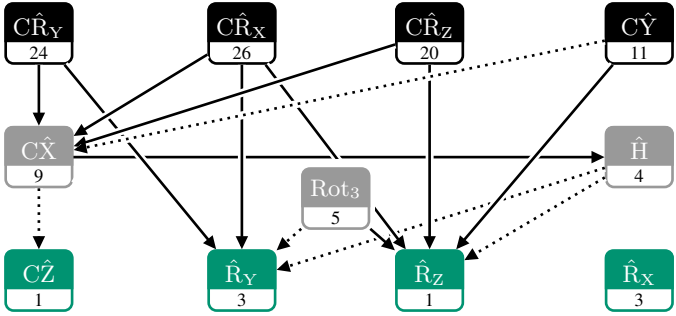

In the context of this work, we choose the universal gate set $\{\RX, \RY, \RZ, \CZ\}$ as our basis gates, on which we build other single- and two-qubit gates.
While these basis gates are directly described by the pulses introduced in \autoref{sec:pulse-shapes}, \autoref{fig:gates_graph} depicts the corresponding dependence of higher-level gates.
Here, the number below each node represents the overall number of pulse parameters available to this specific gate.
As an example, the $\CRY$ gate is composed of two $\RY$ gates and two $\CX$ gates which in turn consist of two $\hadamard$ and one $\CZ$ gate.
There may be several repetitions of a basis gate within a composed gate, which can introduce redundant parameters.
Our pulse-level interface, shown in~\autoref{fig:gates_graph}, uses a graph-based structure that allows controlling either the pulse parameters of the basis gates or all the parameters, including redundant ones from the repeated basis gates.
This approach enables more specialised \gls{qoc} routine (\cf~\autoref{sec:qoc}) and potentially enables increasing the fidelity of composed gates.

\subsection{Quantum Optimal Control}
\label{sec:qoc}

\begin{figure}
  \centering
  \input{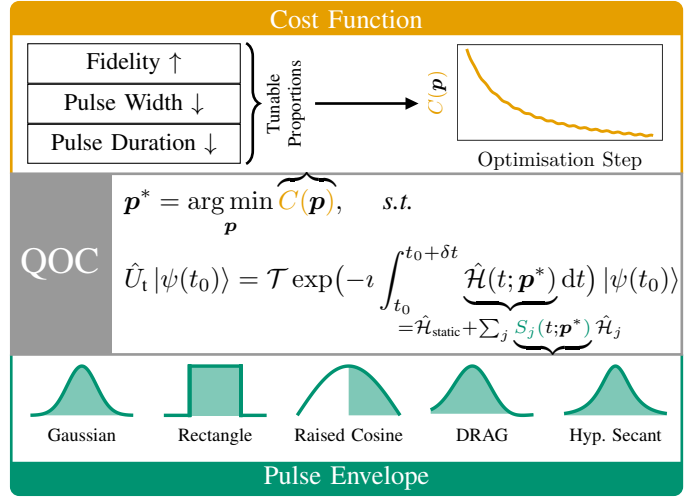}
  \caption{%
    Overview of our \gls{qoc} procedure to find optimal pulse parameters $\bp^*$ by minimising some cost function $C$.
    Resulting pulse parameters determine, for instance, the amplitude, width or duration of the pulse envelope, for which the available possibilities are shown at the bottom.
    For an ideal optimisation, the state after evolution with the constructed pulse gate matches that of the target unitary gate $\hat{U}_\text{t}$.
  }
  \label{fig:qoc_overview}
\end{figure}

We provide a \gls{qoc} routine to learn the pulse parameters of a pulse-based gate.
In \gls{qoc}, overviewed in \autoref{fig:qoc_overview}, we optimise a multi-objective cost function using Adam~\cite{kingma_adam_2017} with weight decay regularisation and a cosine decay learning rate scheduler~\cite{loshchilov_sgdr_2017} with a warm-up phase.
This optimisation is embedded into a multi-stage procedure:
In the first stage, a grid scan is performed to evaluate parameter candidates, each refined through a few gradient steps.
The best-performing candidate is then used to initialise the second stage, which consists of multiple independent optimisation restarts.
The first restart begins from the selected candidate, while subsequent restarts apply Gaussian-distributed random perturbations to the parameter vector.

Maximising the fidelity $\lvert\braket{\psi}{\phi}\rvert^2$ (absolute and phase difference) between the standard unitary state $\ket{\psi}$ and the pulse-based state $\ket{\phi}$ is the primary objective.
Furthermore, minimising pulse duration keeps the sequence as short as possible, which is advantageous due to \tone and \ttwo coherence time constraints.
In addition to these objectives, our \gls{qoc} routine can be extended with custom cost functions accounting for factors like Gaussian pulse width or spectral density.
The state after applying a given pulse sequence can typically only be obtained by measuring multiple shots of a quantum system in an actual hardware environment.
However, as a proof of concept, we compute the pulse-based state by numerically simulating the pulse sequences and compare it to the standard unitary state.

We conducted an optimisation run on the basis gate set described in~\autoref{sec:pulse_gates}.
For the optimisation, we use a mixed-objective cost function consisting of the three parts shown at the top of \autoref{fig:qoc_overview}, where we weighted the pulse width and duration with $5$ and $15$ ($\times 10^{-9}$), respectively. 
\pagebreak

The absolute and phase differences are weighted equally in this optimisation with the remainder to sum up to one (\ie, $1 - 2 \times 10^{-8}$).
For the rotational basis gates $\RX$, $\RY$ and $\RZ$ we calculate the average fidelity over $20$ different rotation angles uniformly distributed in $[0, 2\pi]$.
The $\CZ$ entangling gate is evaluated in combination with leading $\RY$ and $\hadamard$ unitary gates applied on the control and target qubits, respectively, where the $\RY$ gate is swept similarly to the single-qubit optimisation.

\autoref{tab:gate_fidelities_transposed} lists mean results for the absolute ($\Delta \vert \cdot \vert$) and phase ($\Delta \angle \cdot$) differences of the basis gates after an optimisation over the parameter samples.
As the optimisation also minimises the pulse width, which contradicts the fidelity, the results converge to exactly this weighting between the two objectives.

\begin{table}[htb]
  \centering
  \caption{Mean gate infidelity (with respective standard deviations) over the variational parameter domain for all of the optimised pulse gates in \gls{qoc}, compared to unitary gates.}
  \label{tab:gate_fidelities_transposed}
  \footnotesize
  \renewcommand{\arraystretch}{0.95}

  \begin{tblr}{
    width=\linewidth,
    colspec={l *{4}{X[c]}},
    cells={mode=math},
    column{1}={mode=text},
    colsep=2pt,
    leftsep=0pt,
    rightsep=0pt,
    row{3}={valign=m},
  }
    \toprule
    Gate                       & \RX & \RY & \RZ & \CZ \\
    \midrule
    $\Delta \vert \cdot \vert$ &
    \makecell{(1.6\pm1.2)                              \\\times10^{-13}} &
    \makecell{(3.6\pm2.9)                              \\\times10^{-14}} &
    \makecell{(6.3\pm10)                               \\\times10^{-17}} &
    \makecell{(1.6\pm1.2)                              \\\times10^{-8}} \\
    $\Delta \angle \cdot$      &
    \makecell{(0.0\pm0)                                \\\times10^{-13}} &
    \makecell{(0.0\pm0)                                \\\times10^{-14}} &
    \makecell{(0.0\pm0)                                \\\times10^{-17}} &
    \makecell{(17\pm6.1)                               \\\times10^{-5}} \\
    \bottomrule
  \end{tblr}
\end{table}

\begin{figure}[htb]
  \centering
  \includegraphics[width=\columnwidth]{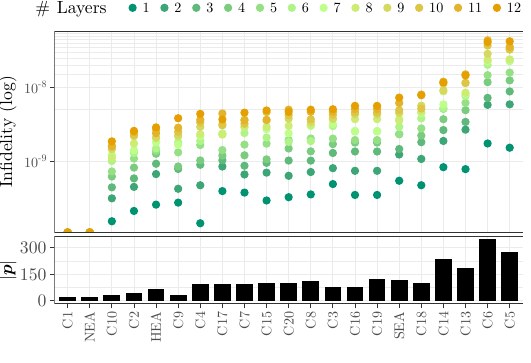}
  \caption{(top) Infidelities for all available \ase, ordered by the mean infidelity over all numbers of layers. (bottom) the number of pulse parameters for each \as. \Ase of the format C* are taken from Ref.~\cite{sim_expressibility_2019}. The additional \ase, NEA, SEA and HEA refer to a \emph{non-entangling}, \emph{strongly-entangling} and \emph{hardware-efficient} \as, respectively. Note that the infidelity is zero for all layers of C1 and NEA.}
  \label{fig:circ_fidelities}
\end{figure}

While the infidelity might appear sufficiently small at first sight, it is important to note that the error accumulates with the size of the circuit.
We demonstrate this effect in~\autoref{fig:circ_fidelities}, where we show the infidelity (average of absolute and phase differences) for all available \ase with circuit depths up to twelve variational layers.
Comparing the results with the circuit structure, it becomes apparent that the number of non-basis gates in a circuit correlates with the infidelity.
Notably, even the highest infidelity in \autoref{fig:circ_fidelities} is orders of magnitude below practically achieved infidelities on real devices, which currently record $10^{-7}$~\cite{smith_single_2025} for single qubit gate fidelities (\cf \autoref{tab:gate_fidelities_transposed}).
Our \gls{qoc} procedure is therefore sufficiently precise in ideal and non-noisy environments.

\section{Metrics}
\label{sec:metrics}

In this section, we describe the analysis metrics shown at the bottom of \autoref{fig:software_overview}, discuss the corresponding implementation details, and show some proof-of-concept results for each metric.

\subsection{Fourier Spectral Properties}
\label{sec:fourier-metrics}

Calculating the Fourier spectrum of \glspl{qfm} is a central tool for analysing the capabilities of different \ase and data encodings~\cite{wiedmann_fourier_2024,mhiri_constrained_2024,sweke_potential_2025,strobl_fourier_2025,franz_out_2025} (\cf \autoref{sec:preliminaries}).
Therefore, computing Fourier coefficients, frequencies and other metrics that derive from them are a key feature of the \qmlessentials package, which we describe in the following.

\subsubsection{Coefficients}
\label{sec:coefficients}
Fourier coefficients can be obtained with two different approaches in the \qmlessentials package:
Either an analytical expression for the coefficients can be derived for a model, or the \gls{fft} is used.

For the former, the analytical method uses an algorithm proposed by \citeauthor{nemkov_fourier_2023}~\cite{nemkov_fourier_2023}, extended in Ref.~\cite{wiedmann_fourier_2024}. 

Here, the expectation value from \autoref{eq:circuitExpr} is expanded in terms of trigonometric polynomials.
However, in its current state, this method is comparatively slow and only implemented for a single input feature, serving as an evaluation of the second, \gls{fft}-based method.
In this second method, the \gls{fft} over a range of input samples yields a numerical approximation of the coefficients based on the expectation value in \autoref{eq:circuitExpr}, given the number of frequencies for a model, for which an upper bound can be derived from the input encoding~\cite{peters_generalization_2023}.

\subsubsection{Fourier Coefficient Correlation}
\label{sec:fcc}

The \gls{fcc} is a metric to quantify the correlation between the intrinsic Fourier coefficients of a \gls{qfm}~\cite{strobl_fourier_2025}.
This is computed numerically by first executing the \gls{fft} to obtain a set of Fourier coefficients while sampling $\btheta \in \Theta$ and then calculating the Pearson correlation coefficient~\cite{freedman_statistics_2007}
\begin{equation}
  r_{\Theta}(\bomega, \bomega')= \frac{\sum_{\btheta\in\Theta}\left(c_{\bomega}(\btheta)-\bar c_{\bomega}\right)\left(c_{\bomega'}(\btheta)-\bar c_{\bomega'}\right)}{\sqrt{\sum_{\btheta\in\Theta}\left(c_{\bomega}(\btheta)-\bar c_{\bomega}\right)^{2} \sum_{\btheta\in\Theta}\left(c_{\bomega'}(\btheta)-\bar c_{\bomega'}\right)^{2}}}
  \label{eq:pearson_correlation}
\end{equation}
between each pair of Fourier coefficients.
Here, $\bar c_{\bomega}$ and $\bar c_{\bomega'}$ represent the average value of the coefficient at frequency $\bomega$ and $\bomega'$, respectively.
This pairwise correlation is termed Fourier fingerprint, and averaging over all pairs results in the \gls{fcc}
\begin{equation}
  \fcc \coloneq \frac{1}{\vert \bOmega \vert} \sum_{\bomega, \bomega' \in \bOmega} \left\vert r_\Theta(\bomega, \bomega') \right\vert.
  \label{eq:fcc}
\end{equation}
Our contribution covers the implementation of both, the Fourier fingerprint and the \gls{fcc} with the default parameters set to replicate the results from Ref.~\cite{strobl_fourier_2025}.

\subsection{Expressibility}
\label{sec:expressibility}

In \qmlessentials, the expressibility is computed by taking the inverse of the \gls{kl}-divergence between states sampled from a given \as and the Haar measure (\cf Apx.~\ref{apx:kl-expr} for details).
Therefore, the expressibility metric introduced in Ref.~\cite{sim_expressibility_2019} evaluates how well an \as can approximate random states in Hilbert space.
However, expressibility does not reflect whether circuits are suitable for learning settings, as functional constraints or input encodings are not considered in this approach.
As shown in Ref.~\cite{strobl_fourier_2025}, the \gls{fcc} addresses this limitation by revealing coefficient dependencies that limit the effective number of learnable frequency components.

\begin{figure}
  \centering
  \includegraphics[width=\columnwidth]{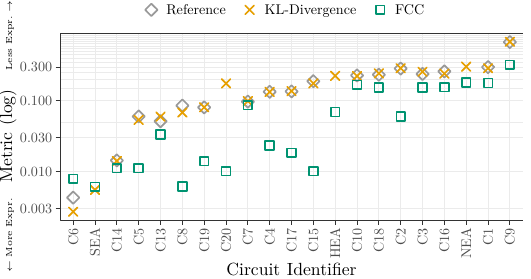}
  \caption{Inverse of the expressibility, that is the \gls{kl}-divergence to Haar states and the \gls{fcc} (\cf Apx.~\ref{apx:kl-expr} and \autoref{sec:fcc}, respectively), on all predefined \ase, which are ordered by the mean across all measures.
    The reference measures are taken from Ref.~\cite{sim_expressibility_2019} and circuit identifiers are described in the caption of \autoref{fig:circ_fidelities}.}
  \label{fig:expressibility-measures}
\end{figure}

The implementation of the expressibility based on the \gls{kl}-divergence from \qmlessentials and our complementary \gls{fcc} implementation allows us to measure and compare both metrics for all available \ase.
In particular, we compute the expressibility for \numprint{10000} sample pairs and the \gls{fcc} for an $\RY$ angle encoding and \numprint{10000} samples of uniformly randomly initialised parameters for each four-qubit \as.
A comparison between the \gls{kl}-divergence-based expressibility and FCC is shown in \autoref{fig:expressibility-measures}, also including the reference results for the expressibility from Ref.~\cite{sim_expressibility_2019}.
As shown in the figure, the \gls{kl}-divergence from \qmlessentials and Ref.~\cite{sim_expressibility_2019} match closely, while the \gls{fcc} yields rather different values for most of the \ase.
This indicates that while some \ase do not seem to be very expressive (\eg, C8, C15, or C20), the actual trainability might be better, as the Fourier coefficients are not highly correlated, and vice versa, thus confirming the discrepancy between these metrics as shown in~\cite{strobl_fourier_2025}.

\begin{figure}
  \centering
  \includegraphics[width=\columnwidth]{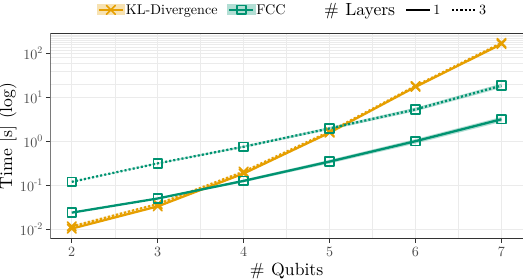}
  \caption{Time required to compute the \gls{kl}-divergence based expressibility and \gls{fcc} for different circuit sizes, using our JAX-based simulator, averaged over ten executions for each configuration.
    The lines and points refer to the mean, and the shaded areas to the standard error of the mean over all available \ase.
  }
  \label{fig:expressibility-benchy}
\end{figure}

To further provide a brief benchmark of our implementation, we measured the runtime required to compute expressibility and \gls{fcc} for an increasing number of qubits.
To this end, we evaluated both metrics for 50 parameter samples (-pairs) on all available \ase with one and three variational layers on the same machine equipped with \texttt{AMD EPYC 9454} processors. 

The results of these benchmark runs, which are shown in \autoref{fig:expressibility-benchy}, demonstrate that the runtime increases exponentially with the number of qubits for both metrics, which is consistent with the exponential growth of the simulated quantum system.
However, steeper curves for the expressibility computation indicate a less favourable scaling behaviour than for the \gls{fcc}.
This aligns with the complexity discussions of Ref.~\cite{strobl_fourier_2025}, where expressibility requires extensive sampling and pairwise fidelity evaluations, while the \gls{fcc} avoids these quadratic comparisons.
Notably, the time for computing expressibility is largely unaffected by the number of layers, while it becomes apparent that the \gls{fcc} scales with the number of intrinsic frequencies of the model and therefore shows an additional computational overhead for the three- versus one-layer configuration.
Overall, these results support the complexity analysis in Ref.~\cite{strobl_fourier_2025} and also demonstrate the fast array computation capabilities of our JAX-based simulator.

\subsection{Entangling Capability}
\label{sec:entanglement}

The entangling capability aims to describe the amount of entanglement a circuit can create.
It can also be understood as a measure of the \emph{quantumness} of a particular circuit, as entanglement is one of the central properties that distinguish a quantum system from classical surrogates.
As there are different ways of quantifying entanglement, there also exist various entangling capability metrics~\cite{vedral_quantifying_1997,plenio_introduction_2007}, of which the \gls{mw} measure~\cite{meyer_global_2002,brennen_observable_2003} and \glspl{bm}~\cite{foulds_controlled_2021,haug_scalable_2023} are included in \qmlessentials.
The definitions and implementation details for these two metrics are summarised in Apx.~\ref{apx:entangling-metrics}.
In addition, we implement the \gls{ef}~\cite{hill_entanglement_1997} and \gls{ce}~\cite{beckey_computable_2021} to also support noisy states.

\subsubsection{Entanglement of Formation}

Following Ref.~\cite{plenio_introduction_2007}, the \gls{ef}~\cite{hill_entanglement_1997} is usually defined as
\begin{equation}
  E_\text{EF}(\ket{\psi}) = \inf \left\{\sum_{i} p_{i} E\left(\ket{\psi_{i}}\bra{\psi_{i}} \right): \rho=\sum_{i} p_{i}\ket{\psi_{i}}\bra{\psi_{i}}\right\},
  \label{eq:entangling_capability_ef}
\end{equation}
and represents the averaged entanglement of a pure state decomposition of a density matrix $\rho$, where the minimum over all possible decompositions then results in the \gls{ef}.
The entanglement for pure states is calculated using the \gls{mw} measure as introduced in~\autoref{eq:entangling_capability_brennen}.
Finding a pure state decomposition, however, is a non-trivial task.
In our implementation, we utilise NumPy's~\cite{harris_array_2020} eigenvalue and eigenstate decomposition of the density matrix $\rho$.
Therefore, this metric is not applicable when using real devices.
Based on the resulting decomposition, we proceed with calculating the entanglement of each eigenstate while weighting it by its eigenvalue as described in the left part of the right-hand side of~\autoref{eq:entangling_capability_ef}.
However, the eigenvalue decomposition is not unique, resulting in several potential decompositions, which can lead to different values for the entangling capability.
Hence, the resulting values of the \gls{ef} represent an upper bound.

\subsubsection{Concentratable Entanglement}

\gls{ce} is a new family of multipartite entanglement measures introduced by \citeauthor{beckey_computable_2021} in \cite{beckey_computable_2021}.
Let $\ket{\psi} \in \mathbb{H}$ be a pure state over $n$ qubits labelled by the set $S = \{1, \ldots, n\}$. For any non-empty subset $s \subseteq S$ of qubit labels, the \emph{Concentratable Entanglement} is defined as
\begin{equation}
  \mathcal{C}_{\ket{\psi}}(s) = 1 - \frac{1}{2^{\lvert s \rvert}} \sum_{\alpha \in 2^s} \Tr[\rho_\alpha^2],
\end{equation}
where $2^s$ denotes the power set of $s$ and $\rho_\alpha$ is the result of tracing out all qubits not in $\alpha$.
CEs can be efficiently computed on a quantum computer using an $n$-qubit parallelised SWAP test \cite{beckey_computable_2021}.
This method requires two copies of the state $\ket{\psi}$ and $n$ additional ancillary qubits, resulting in $3n$ qubits in total.
\citeauthor{beckey_multipartite_2023} show in \cite{beckey_multipartite_2023} that CEs can be approximated from the results of Bell-basis measurements, hence using a circuit without additional ancillas.
CEs, as described, are defined only for pure states. As in the case of the Entanglement of Formation, an entanglement measure for mixed states can be derived from CEs using a convex roof extension:
\begin{equation}
  \mathcal{C_\rho}(s) = \inf \Biggl\{ \sum_i p_i \mathcal{C}_{\ket{\psi_i}}(s) : \rho = \sum_i p_i \ket{\psi_i} \bra{\psi_i} \Biggr\}.
\end{equation}
In general, computing convex roof extension measures presents a hard optimisation problem.
Efficient lower and upper bounds for this value are proven by \citeauthor{beckey_multipartite_2023}~\cite{beckey_multipartite_2023} and \citeauthor{foulds_generalising_2024}~\cite{foulds_generalising_2024}.
Experimentally, however, we observe that computation of the CEs using the $n$-qubit parallelised SWAP test still yields suitable results under the influence of lower levels of noise.

\subsubsection{Experimental Validation}

\begin{figure}
  \centering
  \includegraphics[width=\columnwidth]{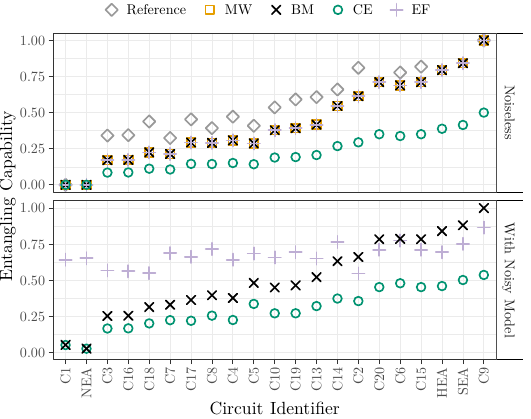}
  \caption{Comparison of different entangling measures on all predefined \ase, which are ordered by the mean entanglement capability across all measures. The reference is taken from Ref.~\cite{sim_expressibility_2019} and circuit identifiers are described in the caption of \autoref{fig:circ_fidelities}.}
  \label{fig:entanglement-measures}
\end{figure}

We compare our implementations against the existing measures in \qmlessentials and the results presented in Ref.~\cite{sim_expressibility_2019}.
Specifically, we evaluate 1000 samples of uniformly randomly initialised parameters for each measure at one layer of each predefined four-qubit \as.
The measures can be separated into those that only work on pure, noiseless states (including the reference~\cite{sim_expressibility_2019} and implemented \gls{mw} measure), and the remaining measures that are also applicable to noisy states; for the latter, we additionally apply a 1\% depolarising noise model. 

The results of this validation are shown in \autoref{fig:entanglement-measures}.
While for pure states, the individual amount and scaling of entangling capabilities differ, the \ase follow a consistent order across all measures.
When applying the noise model, the entangling capability for \gls{bm} and \gls{ce} still follows this order, yet yields slightly higher values than in the noiseless case.
In the noisy setting, the \gls{ef} measure does not follow this increasing pattern in the entangling capability, but reaches higher values for the \emph{lowly} entangled and lower values for the \emph{highly} entangled \ase, according to the other measures.
However, as our \gls{ef} implementation represents a valid upper bound for entangling capability and is based on the same pure state foundation as the \gls{bm}, the \gls{ef} might be a more reliable measure in cases where the \gls{ef} is lower than the \gls{bm}.

This comparison highlights that it can be useful to consider different metrics for differing comparison baselines.
For instance, the \gls{bm} or \gls{ce} measures can be suitable for showing entangling capability differences across \ase at one specific noise level.
However, when conducting comparisons across different noise levels, it might be useful to incorporate entanglement measures such as the \gls{ef}~\cite{franz_out_2025}.

\section{Conclusion and Outlook}
\label{sec:concl}

The software library introduced in this work provides a comprehensive environment for studying \gls{qml} models from both a Fourier-analytic and hardware-aware perspective.
By combining modular \as construction, pulse-level modelling, Fourier-spectral diagnostics, entanglement analysis, and a differentiable simulation backend, our software enables a reproducible workflow for a broad range of controlled numerical experiments within a single package. 
\pagebreak

This paves the way to investigate architectural \as choices and physical constraints in the context of \gls{qml}.
At the same time, we acknowledge that significant computational costs are associated with training \gls{qml} models, particularly in pulse-level and noise-aware settings.
While our framework incorporates performance optimisations to mitigate computational overheads, the primary contribution of this work is scientific rather than immediately application-driven.

In particular, we view the framework as a tool for the systematic and reproducible investigation of pulse-level learning with \glspl{qfm}.
While both perspectives have independently received considerable attention in the \gls{qml} literature, with pulse-level approaches for improving hardware control and Fourier analysis for understanding the functional structure of circuits, to our best knowledge, our software is the first to allow for their combined investigation.
With the foundation provided in this work, we leave experiments and analysis to connect the physical implementation of \gls{qml} models with their spectral properties to future work.

Our framework further provides the basis for a variety of systematic studies.
For instance, it can be investigated to a greater extent how specific data encoding strategies and \ase, with their intrinsic entangling capability and expressibility, influence trainability and set quantum models apart from classical \gls{ml} models.
Additionally, the differentiable pulse-level backend enables experiments for the direct optimisation of pulse parameters in certain learning tasks, as well as the exploration of error mitigation strategies.
Notably, our implementation allows combining pulse-level computation with any of the available noise models, therefore enabling a hardware-aware study of \gls{qml} models.
In future work, this may include adding hardware-aware system dynamics, such as damping due to hardware limitations.
Ultimately, we aim to provide a comprehensive toolbox for reproducible \gls{qml} research.

\section*{Acknowledgment}

We thank \blackout{Gabriel Mejía Ruiz} for valuable discussions and comments to this work.
\blackout{MS}, \blackout{EK} and \blackout{AS} acknowledge support by \blackout{the state of Baden-W\"urttemberg through bwHPC}.
\blackout{LS} acknowledges the support by the Doctoral School \blackout{Karlsruhe School of Elementary and Astroparticle Physics: Science and Technology}.
\blackout{MF} and \blackout{WM} acknowledge support by \blackout{German Research Foundation}, grant \blackout{MA 9739/1-1}, \blackout{the German Federal Ministry of Research, Technology and Space (BMFTR)}, funding program \blackout{\enquote{Research Program Quantum Systems}}, grant number \blackout{13N17387} and the \blackout{High-Tech Agenda of the Free State of Bavaria}.

\appendices

\section{Expressibility Metric}
\label{apx:expressibility-metrics}
\label{apx:kl-expr}

This section provides the definitions for the expressibility metrics implemented in \qmlessentials.
The calculation is based on the \gls{kl}-divergence between the distributions
\begin{enumerate*}[label=(\arabic*)]
  \item obtained by sampling from the Haar integral $\int_{\text {Haar }}\text{d} \psi (\ket{\psi}\bra{\psi})^{\otimes t} $ of a state $t$-design, and
  \item obtained from the model $\int_{\btheta}\text{d}\btheta \left(\ket{\psi_{\btheta}}\bra{\psi_{\btheta}}\right)^{\otimes t}$~\cite{kullback_information_1951,sim_expressibility_2019}:
\end{enumerate*}
\begin{equation}
  D_{\mathrm{KL}}\left(P_{\text{Model}}(F ; \btheta) \| P_{\text {Haar }}(F)\right).
  \label{eq:kl_divergence}
\end{equation}
Here, the fidelity $F \coloneq \lvert\braket{\psi_1}{\psi_2}\rvert^2$ is based on the state overlap of pairs of sampled states $(\ket{\psi_1}, \ket{\psi_2})$, and the distribution is then $P\left(F\right)$.

This metric yields zero if $P_{\text{Model}}(F ; \btheta) = P_{\text {Haar }}(F)$, meaning the states sampled from the \gls{qfm} are Haar distributed.
For the least expressive case, that is, the empty circuit $U=\mathds{1}$, the KL divergence becomes $\ln(n_\text{bins})$ where $n_\text{bins}$ describes the number of bins that are used for discretising the probability distribution using a histogram.
Expressibility is thus regarded as the inverse \gls{kl}-divergence.

\section{Entangling Capability Metrics}
\label{apx:entangling-metrics}

This section provides the definitions for the implemented entangling capability metrics in \qmlessentials.

\subsection{Meyer-Wallach Measure}
The \gls{mw} entangling capability~\cite{meyer_global_2002,brennen_observable_2003} of a state $\ket{\psi}$ is defined as
\begin{equation}
  E(\ket{\psi})=2\left(1-\frac{1}{n} \sum_{k=0}^{n-1} \Tr\left[\rho_{k}^{2}\right]\right),
  \label{eq:entangling_capability_brennen}
\end{equation}
based on partial density matrices $\rho_k$, where $k$ denotes the subsystem (\ie, qubit index).

This metric has the property that if $\Tr\left[\rho_{j}^{2}\right]=1 \quad \forall j$, implying $E=0$, $\ket{\psi}$ is a product state whereas $E=1$ iff $\Tr\left[\rho_{k}^{2}\right]=1 / 2 \quad \forall k$ meaning the state is maximally mixed. Notably, this metric is restricted to pure states and cannot be used for mixed states as they occur in decoherent noisy circuits.

\subsection{Bell Measurement}
The \gls{bm}~\cite{foulds_controlled_2021,haug_scalable_2023} provides an alternative way to compute $\Tr\left[\rho_{k}^{2}\right]$ in \autoref{eq:entangling_capability_brennen}.
Here, a circuit is considered, where the state of interest is vertically prepared twice, thereby requiring $n$ additional ancillary qubits, and an inverse Bell-state is applied onto each pair of qubits between the two subsystems.

As shown in Ref.~\cite{haug_scalable_2023}, the squared trace of the density matrix is linearly dependent on the probability of measuring the parity between each of the qubits in the individual subsystems:
\begin{equation}
  \Tr\left[\rho_{k}^{2}\right]=1-2 \cdot P_{\text {odd}, k},
\end{equation}
where $P_{\text {odd}, k}$ is the probability of odd, non-zero parity in the outcomes of the $k$th qubit on each copy.
Inserting this in \autoref{eq:entangling_capability_brennen}, yields the same estimate of the entangling capability as when using the \gls{mw} metric, yet uses a physical observable measurement applicable to real quantum systems.
However, when using mixed states, impurities from the interaction with the environment may overestimate the entangling capability~\cite{foulds_generalising_2024}.
Therefore, the resulting values of the \gls{bm} for mixed states rather present an upper bound for the entangling capability.

\printbibliography

\end{document}